\documentclass[aps,reprint,showpacs,preprintnumbers,amsmath,amssymb,nofootinbib,superscriptaddress]{revtex4-1}

\usepackage{epsf,color,amsmath,amsthm,amssymb,bm}
\usepackage[dvips]{graphicx}
\usepackage{textcomp}
\usepackage{enumerate}
\usepackage{epsfig}
\usepackage{subfigure}
\usepackage{longtable}
\usepackage{appendix}
\usepackage{comment}
\usepackage{natbib}
\usepackage{accents}
\usepackage[normalem]{ulem}
\usepackage{cancel}
\usepackage{longtable}

\newcommand{\be}{\begin{equation}}
\newcommand{\ee}{\end{equation}}
\newcommand{\bea}{\begin{eqnarray}}
\newcommand{\eea}{\end{eqnarray}}
\newcommand{\bc}{\begin{center}}
\newcommand{\ec}{\end{center}}

\newcommand{\eps}{\varepsilon}

\newcommand{\mrm}{\mathrm}

\begin{document}
\title{Monte Carlo simulation of electrostatic interactions in inhomogeneous dielectric media: Correct sampling for the local lattice simulation algorithm}
\author{Xiaozheng Duan}
\affiliation{State Key Laboratory of Polymer Physics and Chemistry, Changchun Institute of Applied Chemistry, Chinese Academy of Sciences, Changchun, Jilin 130022, China}
\author{Issei Nakamura}
\email{inakamur@mtu.edu} 
\affiliation{State Key Laboratory of Polymer Physics and Chemistry, Changchun Institute of Applied Chemistry, Chinese Academy of Sciences, Changchun, Jilin 130022, China}
\affiliation{Department of Physics, Michigan Technological University, Houghton, Michigan 49931, United States}
\author{Zhen-Gang Wang}
\email{zgw@caltech.edu}
\affiliation{Division of Chemistry and Chemical Engineering, California Institute of Technology, Pasadena, California 91125, United States}

\date{\today}

\begin{abstract}
We present a lattice Monte Carlo algorithm based on the one originally proposed by Maggs and Rossetto for simulating electrostatic interactions in inhomogeneous dielectric media.  The original algorithm is known to produce attractive interactions between particles of the same dielectric constant in the medium of different dielectric constant.  We demonstrate that such interactions are spurious, caused by incorrectly biased statistical weight arising from particle motion during the Monte Carlo moves. We propose a simple parallel tempering algorithm that corrects this unphysical bias.  The efficacy of our algorithm is tested on a simple binary mixture and on an uncharged polymer in a solvent, and applied to salt-doped polymer solutions.
\end{abstract}

\maketitle
There is a persistent need for efficient computational methods for modeling dielectric responses of materials and electrostatic interactions in dielectric media.  In spite of many advances in methodology development over the past few decades, the computational effort for directly simulating charge systems remains considerably demanding \cite{cisneros14}. Specifically, accurate treatment of systems with spatially varying dielectric permittivity still poses a substantial challenge in both implicit \cite{maggs06,shen02,barros14} and explicit \cite{papazyan97,levy98,sagui99} solvent models.  Continued progress in the study of electrostatic interactions in diverse systems and phenomena requires a fundamental breakthrough in the simulation techniques capable of efficiently treating dielectric inhomogeneities \cite{messina09,jadhao12,barros14}.    

Maggs and Rossetto proposed a local lattice Monte Carlo (MC) algorithm based on solving Gauss's law \cite{maggs02}, which scales with the particle number $N$ as $O(N)$.  While the algorithm has proven to be strikingly efficient in computational performance \cite{maggs02,rottler04,duncan05,rottler11}, the original method is known to produce attractive interactions between uncharged bodies of one dielectric permittivity in the medium of different dielectric permittivity.   Duncan et al. \cite{duncan06} argued that such attractive force is spurious, likely caused by an inaccurate statistical sampling; yet the nature and origin of the attractive force remain controversial \cite{rottler11}.

The local simulation algorithm to solve Gauss's law can be articulated by introducing the identity \cite{nakamura13_SoftMatter}
\bea
\label{e:identity}
\exp(-U_\mrm{el}[\vec{D}]) &=&  Z_{\mrm{fluc}}^{-1} (\{r_\mrm{i}\})  \int {\mathcal D}\vec{A} \biggl\{\prod_{\vec{r}} \delta [\mrm{div}\vec{A}(\vec{r})-\rho(\vec{r})]\biggr\} \nonumber\\
  &\times& \exp(-U_\mrm{el}[\vec{A}]).
\eea
Here, $U_\mrm{el}[\vec{D}] = \int d\vec{r}\, \frac{[\vec{D}(\vec{r})]^2}{2\eps_0\eps(\vec{r})}$ is the electrostatic energy, with $\vec{D}(\vec{r})$ being the electric displacement and $\rho(\vec{r})$ the local charge density. $\eps_0$ and $\eps(\vec{r})$ are the vacuum permittivity and local dielectric function, respectively. The particle position is denoted collectively by $\vec{r}_i$ and includes all charged and uncharged species. $Z_{\mrm{fluc}}=\int {\mathcal D}\vec{A}_t \{\prod_r \delta [\mrm{div}\vec{A}_t]\}\exp(-U_\mrm{el}[\vec{A_t}])$  where $\vec{A}_t(\vec{r})$= $\vec{A}(\vec{r}) - \vec{D}(\vec{r})$ is the transverse vector field and div$\vec{D}(\vec{r}) = \rho(\vec{r})$.  Using Eq.\,(\ref{e:identity}), the statistical average of any property $S$ can be written as \cite{nakamura13_SoftMatter}:
\bea
\label{e:mc_formula}
\langle S\rangle &=& \int \prod_{i} d\vec{r_i} Z_{\mrm{fluc}}^{-1} (\{r_\mrm{i}\}) {\mathcal D} \vec{A} \biggl\{\prod_{\vec{r}} \delta [\mrm{div}\vec{A}(\vec{r})-\rho(\vec{r})]\biggr\} S \nonumber\\
&\times& \exp\{-U_\mrm{el}[\vec{A}]\}/\int \prod_{i} d\vec{r_i} Z_{\mrm{fluc}}^{-1} (\{r_\mrm{i}\}) {\mathcal D} \vec{A} \nonumber\\
&\times&\biggl\{\prod_{\vec{r}} \delta [\mrm{div}\vec{A}(\vec{r})-\rho(\vec{r})]\biggr\} \exp\{-U_\mrm{el}[\vec{A}]\}.
\eea
Taking $S$ to be $\vec{A} (\vec{r})$, the electric displacement is calculated to be 
\bea
\label{e:D_A}
\vec{D}(\vec{r}) =\langle\vec{A}(\vec{r})\rangle.
\eea
The local lattice MC algorithm involves statistical sampling of both the position of the species $\vec{r}_i$ and the auxiliary field variable $\vec{A}(\vec{r})$ in accordance with Eq.\,(\ref{e:mc_formula}).

The auxiliary field $\vec{A}(\vec{r})$ is not to be confused with the {\it physical} electric displacement field $\vec{D}(\vec{r})$. Indeed, the electric displacement field satisfies both rot$\vec{D}(\vec{r}) = 0$ and div$\vec{D}(\vec{r}) = \rho(\vec{r})$ in the absence of a time-dependent magnetic field, whereas div$\vec{A}(\vec{r}) = \rho(\vec{r})$ is the only constraint for the auxiliary field. Since $\vec{A}(\vec{r})$ can be written as $\vec{A}(\vec{r}) = \vec{D}(\vec{r}) + \text{rot}\,\vec{Q}(\vec{r})$, where $\vec{Q}(\vec{r})$ is an arbitrary vector, the functional integration over the transverse vector can be cast into $Z_{\mrm{fluc}} = \int {\mathcal D} \vec{Q} \exp\{-[\text{rot}\,\vec{Q}(\vec{r})]^2 /[2\eps_0\eps(\vec{r})] \}$. 

For homogeneous dielectric media, $Z_{\mrm{fluc}}$ is constant and therefore thermodynamically inconsequential \cite{maggs02}.  However, for systems where the dielectric function depends on the particle positions of the species, no simple treatment of $Z_{\mrm{fluc}}$ exists.  Therefore, $Z_{\mrm{fluc}}$ is commonly neglected.  Such a treatment is known to result in the attractive interactions between uncharged species, which was rationalized as arising from the dispersion interaction \cite{maggs04}.  However, dispersion interaction has its origin in quantum mechanics \cite{TF937330008B} whereas the local simulation algorithm is just a convenient mathematical scheme to solve Gauss's law (or equivalently Poisson's equation).   Accordingly, the scheme should only account for purely classical effects when the Hamiltonian includes only classical degrees of freedom and interactions.  Therefore, when there is no charge in the system, we expect $D(\vec{r}) = 0$ from div$\vec{D}(\vec{r}) = 0$ and rot$\vec{D}(\vec{r}) = 0$, regardless of any dielectric inhomogeneity.  Thus, the energy landscape of the system given by $U_\mrm{el}[\vec{D}]$ should be uniform and independent of particle configurations, indicating no net electrostatic interaction between the uncharged species. In this Letter, we first demonstrate that, indeed, the local simulation algorithm does not produce any force in the systems containing no charge.  We then present a simple, general method to avoid the undesirable spurious interactions in the local lattice MC simulations via parallel tempering.

We first consider the statistical average of the auxiliary field $\langle\vec{A}(\vec{r})\rangle$ for a system of uncharged species with their positions fixed. Eq.\,(\ref{e:mc_formula}) then becomes
\bea
\label{e:avg_A}
\langle \vec{A}(\vec{r}) \rangle &=& \vec{D}(\vec{r}) = \int {\mathcal D} \vec{A} \biggl\{\prod_{\vec{r}} \delta [\mrm{div}\vec{A}(\vec{r})-\rho(\vec{r})]\biggr\} \vec{A}(\vec{r}) \nonumber\\
&\times& \exp\{-U_\mrm{el}[\vec{A}]\}/\int  {\mathcal D} \vec{A} \nonumber\\
&\times&\biggl\{\prod_{\vec{r}} \delta [\mrm{div}\vec{A}(\vec{r})-\rho(\vec{r})]\biggr\} \exp\{-U_\mrm{el}[\vec{A}]\}.
\eea
Note that the factor $Z_{\mrm{fluc}}$ does not appear in this equation because the positions of the species are fixed.   When the particles are allowed to move, the value of $Z_{\mrm{fluc}}$ changes with the particle configuration, and does not cancel out in performing the statistical average Eq.\,(\ref{e:mc_formula}).  Neglecting $Z_{\mrm{fluc}}$ therefore generates unphysical bias during the simulation that involves particle moves.

\begin{figure}[htp!]
\includegraphics[scale=0.28,angle=0]{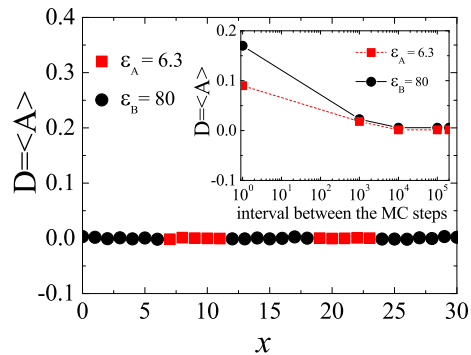}
\caption{
(a) Electric displacement field $\vec{D}(\vec{r})$ at the position $x$ in the simulation box.}
\label{fig:Fig1}
\end{figure}

\begin{figure}[htp!]
\includegraphics[scale=0.28,angle=0]{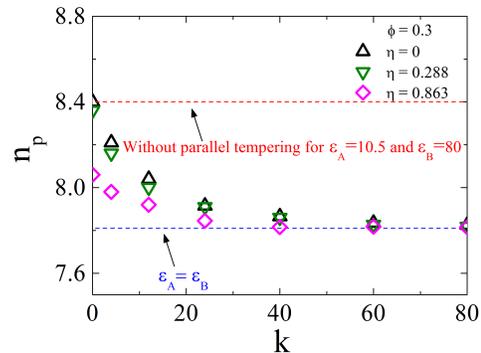}
\caption{The number of neighboring particles $n_{p}$ for particle A as a function of  the number of updates of the auxiliary field $\vec{A}(\vec{r})$. Colored symbols correspond to the results for our parallel tempering method.}
\label{fig:Fig2}
\end{figure}

\begin{figure}[htp!]
\includegraphics[scale=0.28,angle=0]{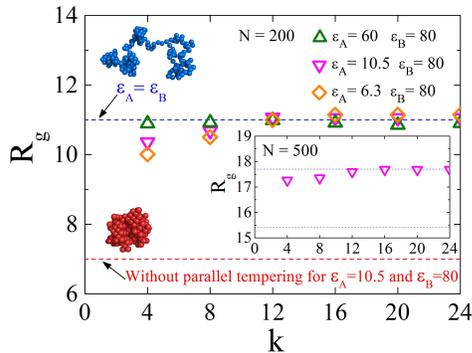}
\caption{The radius of gyration $R_g$ for a single uncharged polymer as a function of the number of updates of the auxiliary field $\vec{A}(\vec{r})$.}
\label{fig:Fig3}
\end{figure}

\begin{figure}[htp!]
\includegraphics[scale=0.17,angle=0]{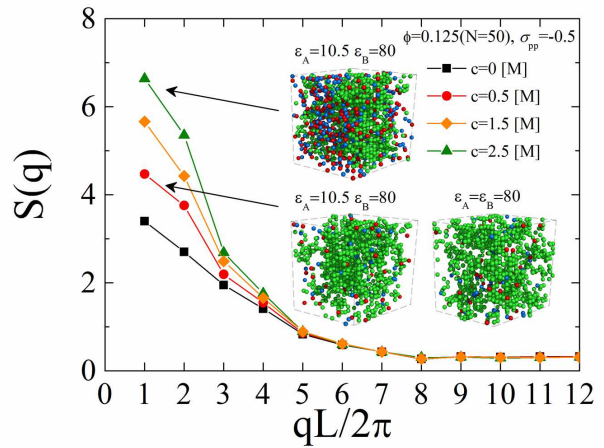}
\caption{Structure factor vs. the wave number.}
\label{fig:Fig4}
\end{figure}

To model the dielectric inhomogeneity, we adopt the same scheme for liquid mixtures as proposed by Maggs \cite{maggs04}: we assign the dielectric constants $\eps_A$ and $\eps_B$ to the lattice grids \cite{maggs02,nakamura13_SoftMatter} occupied by lower-dielectric and higher-dielectric components, respectively.  We perform Metropolis sampling using the Boltzmann weight with energy $A_i^2/(2\eps_0\eps_i)$ on the i-th lattice site. The current of $A_i$ through a surface of the cube that bisects the lattice bond connecting $i$ and $j$ is expressed as $b_{ij}$.  A group of the lattice variables $b_{ij}$ are sequentially updated on a lattice plaquette \cite{nakamura13_SoftMatter,duan15} using the Boltzmann weight subject to the lattice version of Gauss's law $\sum_j b_{ij} = q_i$.  Here, $q_i$ is the charge placed on the lattice grid. The sizes of the simulation box are set to $L^3 = (30u)^3$ for Fig.\,\ref{fig:Fig1} and Fig.\,\ref{fig:Fig3}, $L^3 = (10u)^3$ for Fig.\,\ref{fig:Fig2}, and $L^3 = (20u)^3$  for Fig.\,\ref{fig:Fig4}, where the lattice unit $u = 2.8$ \AA\,. Periodic boundary conditions are used in all directions.  Interested readers are referred to a more complete explanation of the update scheme for the auxiliary field $\vec{A}(\vec{r})$ in Refs.\,\cite{nakamura13_SoftMatter} and \cite{duan15}.

In Fig.\,\ref{fig:Fig1}, we consider a {\it fixed} stripe pattern of the two dielectric species with no charge. The uncharged components with $\eps_A = 6.3$ and $\eps_B = 80$ are distributed periodically.   We performed statistical analysis by averaging on the order of 3000 samples; good statistical convergence was observed on the order of 100 samples.  Note that the simulation for Eq.\,(\ref{e:avg_A})  results in $\vec{D}(\vec{r}) = 0$ throughout the simulation box, indicating no electrostatic interaction between the species. We also considered a non-periodic pattern but again obtained $\vec{D}(\vec{r}) = 0$. This implies that any spurious force between uncharged species when particles are allowed to move must be due to incorrect sampling as a result of neglecting $Z_{\mrm{fluc}}$ in the MC moves.  Furthermore, the inset shows the statistical average of the auxiliary field $\vec{A}(\vec{r})$ as a function of the MC steps between consecutive measurements in equilibrium. We find that a minimum of $10^4$ MC steps are required to obtain statistically independent samples to reach the correct average of $\vec{A}(\vec{r})$. This result also suggests that at least $10^4$ MC steps are required to equilibrate the MC simulation.  Thus, combining a straightforward implementation of the local simulation algorithm at fixed particle positions that correctly samples the electrostatic interactions with other simulation methods for moving the particles such as molecular dynamics, poses formidable computational challenges.  Existing hybrid methods essentially take $\vec{A}(\vec{r})$ to be $\vec{D}(\vec{r})$\cite{rottler04,pasichnyk04,rottler11,fahrenberger14,fahrenberger15}, which is tantamount to neglecting the factor $Z_{\mrm{fluc}}$.

We propose a scheme that uses parallel tempering to avoid the biasing errors introduced through the neglect of $Z_{\mrm{fluc}}$ in the conventional Metropolis implementation of the lattice MC moves.  We start by recalling the key ideas of the parallel tempering method as relevant to our current problem.  We consider states $a$, $b$, $c$ and $d$ with the energies $E_a$, $E_b$, $E_c$, and $E_d$ that correspond to different particle configurations and temperatures $T_0$, $T_1$, $T_2$, and $T_3$. The corresponding canonical probability distribution for the state $\alpha$ at temperature $T_i$ is  $p_i(E_{\alpha}) = Z_i^{-1} \exp \bigl(\frac{-E_{\alpha}}{T_i}\bigr)$, where $Z_i$ is the normalization factor for temperature $T_i$.  For simplicity, we set the Boltzmann constant $k_b$ to 1. We consider parallel tempering with the swaps of the states $a$ and $d$ between $T_1$ and $T_2$, and $b$ and $c$ between $T_0$ and $T_3$. The Boltzmann weight for these two simultaneous swaps is $p_\mrm{swap} = \exp{\bigl[-\frac{(E_a - E_d)}{T_2}\bigr]}  \exp{\bigl[-\frac{(E_d - E_a)}{T_1}\bigr]}  \exp{\bigl[-\frac{(E_b - E_c)}{T_0}\bigr]}$ $\exp{\bigl[-\frac{(E_c - E_b)}{T_3}\bigr]}$.

For subsequent discussions, we consider the special limiting case with $T_2 \rightarrow \infty$ and $T_3 \rightarrow \infty$. The Boltzmann weight for the swaps then becomes
\bea
\label{e:swap_eff}
p_\mrm{swap}  &=& \exp{\biggl[-\dfrac{(E_d - E_a)}{T_1}\biggr]}\exp{\biggl[-\dfrac{(E_b - E_c)}{T_0}\biggr]}.
\eea 
We will use Eq.\,(\ref{e:swap_eff}) to deal with the contributions from the functional integration over the transverse vector in the local simulation method.

We write the dielectric constants for species A and B as $\eps_B = \eps_A + \eta \Delta \eps$.  The real system has $\eta=1$ but for parallel tempering, we allow the dielectric contrast parameter $\eta$ to take any value between $0$ and $1$.  Let $E(i, \eta, \vec{A})$ denote the energy of a microstate with particle configuration $i$, dielectric contrast $\eta$, and auxiliary field $\vec{A}$.  In the local simulation algorithm, Eq.\,(\ref{e:mc_formula}) defines $E(i, \eta, \vec{A})$ as $E(i, \eta, \vec{A}) = \ln Z_\mrm{fluc}(i, \eta) + \int d\vec{r} \vec{A}(\vec{r})^2 / [2\eps(i, \eta)]$.  For simplicity, we scale the field $\vec{A}$ so that $\eps_0 = 1$.

To make use of Eq.\,(\ref{e:swap_eff}), we let the state be represented by the particle configuration $i$ and auxiliary field $\vec{A}$, and let $\eta$ play the role of the temperature $T$.  We consider the Boltzmann weight for the following swaps:
\bea
\label{e:swap_local}
p_\mrm{swap}  &=& \exp{\Bigl\{-[E(j, 1, \vec{A}^{\prime}) - E_ (i, 1, \vec{A})]\Bigr\}}\nonumber\\
&\times& \exp{\Bigl\{-[E(i, \eta, \vec{B}) - E(j, \eta, \vec{C})]\Bigr\}} \nonumber\\
&=& \exp{\Big[E (j, \eta, \vec{C}) - E(j, 1, \vec{A}^{\prime})\Bigr]}\nonumber\\
&\times& \exp{\Bigl[E(i, 1, \vec{A}) - E(i, \eta,  \vec{B})\Bigr]}.
\eea
Since Eq. \,(\ref{e:swap_local}) holds for arbitrary values of the vector field, we choose $\vec{B}(\vec{r})$ and $\vec{C}(\vec{r})$ such that $E(i, \eta, \vec{B}) = \ln Z_\mrm{fluc}(i,1) + \int d\vec{r}\, \vec{A}(\vec{r})^2 / [2\eps(i, \eta)]$ and $E(j, \eta, \vec{C}) = \ln Z_\mrm{fluc}(j,1) + \int d\vec{r}\, \vec{A}^{\prime}(\vec{r})^2 / [2\eps(i, \eta)]$.  We thus arrive at
\bea
\label{e:swap_final}
p_\mrm{swap}  &=& \exp{\Bigl\{A^{\prime\, 2 }/ [2\eps(j, \eta)] - A^{\prime\, 2} / [2\eps(j,1)]\Bigr\}} \nonumber\\
&\times& \exp{\Bigl\{A^2 / [2\eps(i,1)]  - \Bigl. A^2 / [2\eps(i, \eta)]\Bigr\}} 
\eea
Note that the unknown factor $Z_\mrm{fluc}$ does not appear in Eq.\,(\ref{e:swap_final}). Equation\,(\ref{e:swap_final}) thus allows us to effect the correct MC transitions from the particle configuration $i$ to $j$ and field configuration $\vec{A}$ to $\vec{A}^{\prime}$ in the physical system with $\eta = 1$ without knowledge of $Z_\mrm{fluc}$! 

To ensure good overlap of distribution between the different replicas in parallel tempering, differences in the energies of replicas should be small. To this end, we generate $\vec{A}_n^{\prime}$ from $\vec{A}_n$ by $k$ updates, using the statistical weight $\exp{\{-\sum_{n \in \mrm{plaquette}} [(A_n + a_n)^2 - A_n^2]/[2\eps(i,\eta)]\}}$ with a random number $a_n$. In the context of parallel tempering, the number of replicas increases with increasing $k$.

We first test our parallel tempering method on a binary mixture of particles A and B with $\eps_A = 10.5$ and $\eps_B = 80$. $\Delta \eps$ is therefore $69.5$.  Given the incompressible nature of the lattice model, the volume fractions of particles A and B are denoted by $\phi$ and $1 - \phi$, respectively.  We set $\phi = 0.3$.  We consider A-A pairs defined as when two neighboring particles A are separated by a distance of less than or equal to $\sqrt{3}u$ [Fig.\,\ref{fig:Fig2}]. 

In Fig.\,\ref{fig:Fig2} we show the average number of neighboring A particles around a tagged A particle as a function of the number of updates.  In the absence of any interactions, random mixing gives 7.8, as is confirmed for $\eps_A = \eps_B$. This is shown as the blue dashed line.  The red dashed line for $\eps_A \neq \eps_B$ without parallel tempering is higher than the blue dashed line, indicating effective attraction between the like particles due to incorrect sampling. We note that this spurious attractive force is indeed obtained in Ref.\,\cite{maggs04}. With parallel tempering, for any value of $\eta$ (that is appreciably less than 1), the results converge to the blue dashed line with increasing number of updates of the auxiliary field, $k$.  

As the second example, we consider an uncharged polymer of $N=200$ units with lower dielectric constant $\eps_A = 10.5$ immersed in a higher dielectric medium of $\eps_B = 80$ (mimicking water). To model the polymer conformation, we employ the bond fluctuation model of Shaffer \cite{shaffer94} on a simple cubic lattice with nearest-neighbor distance $u$.  The bond vectors of the consecutive monomers on the polymer are allowed to take the bond lengths, 1, $\sqrt{2}$, and $\sqrt{3}$ in the lattice unit $u$.  An empty site corresponds to the coarse-grained density of water molecules. The details of the model were described in an earlier publication \cite{nakamura13_SoftMatter}.  

Figure \ref{fig:Fig3} shows the radius of gyration $R_g$ for a single uncharged polymer with different values of $\eps_A$.  In the absence of dielectric contrast and other energetic interactions, the polymer is in an athermal solvent, and so the chain is expanded with $R_g$ given by the blue dashed line.  If sampling is done correctly, the polymer conformation should be independent of the dielectric contrast.  When we ignored $Z_\mrm{fluc}$ using conventional MC moves in the case of $\eps_A \neq \eps_B$ (red dashed line), the chain adopts a globular structure due to the spurious attractive interaction.  Our parallel tempering algorithm eliminates this artificial interaction; the colored symbols from our parallel tempering sampling converge to the same $R_g$ as in the case of $\eps_A = \eps_B$ for $k > 8$  .

As a final example, we use our parallel tempering algorithm to study the effect of salts on the miscibility in an uncharged-polymer solution.  We consider a solution of polymers with $N=50$ at volume fraction $\phi=0.125$ and assign the nearest neighbor interaction energy $\sigma_\mrm{pp} = -0.5\, k_b T$ for the monomer-monomer interaction. For simplicity, the interaction energies for the monomer-solvent and solvent-solvent interactions are set to zero.  The dielectric constant of the polymer and solvent are respectively $\eps_A=10.5$ and $\eps_B=80$.  To avoid the volume effects of the salt ions, we allow an anion or cation to occupy the same lattice site as a solvent, but forbid more than one ions from occupying the same site.  To calculate the change in the miscibility, we compute the structure $S(q) = \langle |\sum_n \exp(i\vec{q}\cdot \vec{r}_n)[{\delta_A}^n-{\delta_B}^n-\langle{\delta_A}^n-{\delta_B}^n \rangle] |^2 \rangle/L^3$ \cite{sariban1991spinodal}, where the local concentration variable $\delta_A^n$ (or $\delta_B^n $) is 1 or 0 when the lattice site designated by $\vec{r}_n$ is occupied or unoccupied by a monomer (or solvent).  Due to the periodic boundary condition,  the wavevector is given by $\vec{q} = \frac{2\pi}{L}(n_x, n_y, n_z)$, where $(n_x, n_y, n_z)$ is a group of integers.  We perform spherical averaging to obtain the structure factor as a function of the wavenumber $q$.  As shown in Fig. \ref{fig:Fig4}, $S(q)$ at small $q$ increases with increasing salt concentration $c$.  The results indicate that the polymer (green sphere) and solvent become less miscible with the addition of the salt ions (red and blue spheres), with the salt ions concentrated in the higher-dielectric-constant solvent regions (see the insets). This feature is consistent with predictions from previous coarse-grained mean-field theories \cite{zhen-gang08,nakamura12_prl,nakamura12_soft}.

In summary, we have shown that the original algorithm by Maggs and Rossetto \cite{maggs02} for treating configuration-dependent dielectric inhomogeneities introduces an unphysical bias in the sampling by neglecting the factor $Z_\mrm{fluc}$ from the integration over the transverse auxiliary field.  We have proposed a parallel tempering scheme that effectively corrects for this bias without requiring knowledge of $Z_\mrm{fluc}$.  The extra computational cost associated with generating the auxiliary field in the replica exchange is quite modest and the overall $O(N)$ scaling of the original algorithm is preserved.  Our new algorithm therefore represents an efficient method for simulating a wide range of phenomena involving charge interactions in inhomogeneous, configuration-dependent dielectric media.


This work was supported by the National Natural Science Foundation of China (21474112 and 21404103) and the start-up funds of Michigan Technological University. We are grateful to Computing Center of Jilin Province for their essential support.

%

\end{document}